\documentclass[%
superscriptaddress,
preprint,
showpacs,preprintnumbers,
åÊamsmath,amssymb,
åÊaps,
]{revtex4-1}

\usepackage{graphicx}% Include figure files
\usepackage{bm}% bold math
\usepackage{references}
\usepackage{hyperref} 

\usepackage{equationarray}
\usepackage[normalem]{ulem}
\usepackage[usenames]{xcolor}
\usepackage{booktabs}

\newcommand{\be}{\begin{equation}}
\newcommand{\ee}{\end{equation}}
\newcommand{\ba}{\begin{eqnarray}}
\newcommand{\ea}{\end{eqnarray}}
\newcommand{\bd}{\begin{displaymath}}
\newcommand{\ed}{\end{displaymath}}

\usepackage{ulem} %for strike through i.e newcommand cut

\begin{document}

%\preprint{APS/123-QED}

\title{Neutron stars as probes of extreme energy density matter }% Force line breaks with \\

\author{Madappa Prakash}
\email{prakash@phy.ohiou.edu}
\affiliation{Department of Physics and Astronomy, Ohio University, Athens, OH 45701}

\date{\today}% It is always \today, today,

\begin{abstract}
Neutron stars  have long been regarded as extra-terrestrial laboratories from which we can learn about extreme energy density matter at low temperatures. In this article, I highlight some of the recent advances made in astrophysical observations and related theory. Although the focus is on the much needed information on masses and radii of several individual neutron stars, the need for additional knowledge about the many facets of neutron stars is stressed. 
The extent to which quark matter can be present in neutron stars is summarized with emphasis on the requirement of non-perturbative treatments. 
Some longstanding and  new  questions, answers to which will advance our current status of knowledge, are posed.  \\


\end{abstract}

%Use showkeys class option if keyword
\maketitle

%\tableofcontents 

%\input{1_intro}
\section{INTRODUCTION}
\label{Sec:Intro}

Relativistic gravity is important for phenomena that occur close to the diagonal line $2GM=c^2R$ in the characteristic mass $M$ vs.  characteristic distance $R$ diagram of objects in our Universe ($G$ is Newton's gravitational constant and $c$ is the speed of light) \cite{Hartle}.   Observed neutron star masses lie in the range 1-2~${\rm M}_\odot$, whereas their radii are about 10-15 km so that $GM/c^2R \sim$ 0.1-0.3 (compare this with $G{\rm M}_\odot/c^2R_\odot  \sim 10^{-6}$). This happenstance has made it possible to establish the link between space-time geometry and the internal properties of matter - specifically, the relationship between the pressure and energy density which constitutes the equation of state (EOS) of compact objects - that Einstein's theory of general relativity predicts.
 While the entry of neutron stars in theorists minds dates back to the early 1930's \cite{bz34},  their discovery had to wait till the late 1960's \cite{bell68}. Since then, the confluence of theory, astronomical observations and laboratory experiments has revealed
 that all known forces of nature - strong, weak, electromagnetic and gravitational - play key roles in the formation, evolution, and the composition of neutron stars in which the ultimate energy density of observable cold  matter resides.   
Research on the physics and astrophysics of neutron stars has been the forerunner in the study of extreme energy density physics, spurring investigations of relativistic heavy-ion collisions in which high energy density is investigated at temperatures much higher than those encountered in neutron stars. 

In the last decade, several key astronomical observations of neutron stars have been made. The theoretical interpretations of these  
observations have been pursued vigorously around the world with much insight gained into the structure and thermal evolution of neutron stars. This article highlights  some of these developments, and summarizes selected open issues and challenges.   The topics addressed here reflect predilections that stem from my limited involvement and are by no means exhaustive. The reader is recommended to consult comprehensive reviews, references to some of which are provided in later sections. 

\section{NEUTRON STAR MASSES AND RADII }
\label{Sec:masses}
The two most basic properties of a neutron star are its mass $M$ and radius $R$. These physical traits govern several other observables including~\cite{LP:00,LP:04,LP:07} 
 
\noindent 1. The binding energy B.E. of a neutron star: 
\be
B.E \simeq  (0.6\pm0.05)~\frac{GM^2}{Rc^2}~\left(1-\frac{GM}{Rc^2}\right)^{-1} \,.
\ee
Nearly 99\% this B.E. is carried away by neutrinos emitted during the birth of a neutron star in the aftermath of a type-II (core collapse) supernova explosion.  Currently, several detectors are in place to record these neutrinos should a supernova explosion due to core collapse occur within a detectable distance.

\noindent 2. Minimum spin periods of rotation:
\ba
P_{min}(M_{max}) &=& 0.83~\left(\frac {M_{max}}{\rm M_\odot}\right)^{-1/2} 
\left(\frac{R_{max}}{10~{\rm km}}\right)^{3/2} ~{\rm ms} \,, \nonumber \\
P_{min}(M)&\simeq&(0.96\pm 0.3) ~\left(\frac {M}{\rm M_\odot}\right)^{-1/2} 
\left(\frac{R}{10~{\rm km}}\right)^{3/2} ~{\rm ms} \,, 
\ea
where $M_{max}$ and $R_{max}$ are the non-rotating maximum mass spherical configurations, and the second relation refers to a mass not too close to the maximum mass. Precisely measured periods of $\sim 2000$ radio pulsars are available to date 
(cf. pulsar data bases). 
%(see http://www.atnf.csiro.au/research/pulsar/psrcat \\  and http://www3.mpifr-bonn.mpg.de/old$_$pifr/div/pulsar/data/archive.html for pulsar data bases).  
Sub-milli second pulsars (not yet discovered!) are of much interest as they represent the most compact configurations (largest $M/R$ values) for which effects of general relativity are the largest. 
%\item 

\noindent 3. Moment of Inertia: 
\be
I_{max} = 0.6\times 10^{45} 
\frac {(M_{max}/{\rm M_\odot}) (R_{max}/10~{\rm km})^2}
{1-0.295(M_{max}/{\rm M_\odot})/(R_{max}/10~{\rm km}) }
~~{\rm g~cm^2} \,.
\ee
Accurate pulse timing techniques are needed to measure $I$ (through spin-orbit coupling) in extremely compact binaries \cite{LS:05}. Knowledge of the period $P$ and the moment of inertia $I$ of the same neutron star would break the degeneracy between $M$ and $R$, thus allowing for a simultaneous measurement of $M$ and $R$, a first of its kind in this field.   

%\end{enumerate}  
Reference \cite{LP:07} lists additional observables that are significantly influenced by $M$ and $R$.   
The quantities  $M,~R,~B.E.~,I$, and the surface red-shift $z = (1-2GM/Rc^2)^{-1/2}-1$ can be calculated using
the general relativistic (TOV) equations of stellar structure \cite{TOV1,TOV2} by providing 
the equation of state (EOS) of neutron star matter (the relationship between pressure $p$ and energy density $\epsilon$ at every location in the star) as input. The one-to-one correspondence between the EOS and the observed $M~{\rm vs}~R$ curve can be used to model-independently determine the EOS of neutron star matter \cite{Lind:92} as will be discussed later.  

\subsection{Neutron star masses}
Pulsars in bound binary systems afford the 
most accurate measurements of neutron star masses.
Using pulse-timing techniques \cite{manchester1977}, 
the Keplerian parameters (i) the binary period $P$, (ii) the projection
of the pulsar's semimajor axis on the line of sight $a_p\sin i$ (where
$i$ is the binary inclination angle), (iii) the eccentricity $e$, (iv) the
time,  and (v) longitude of periastron $\omega$ can be   
precisely measured. 
Combining (i) and (ii) yields the the mass function:
\be 
f_p=\left(\frac {2\pi}{P}\right)^2\left({a_p\sin i\over
  c}\right)^3{{\rm~M}_\odot\over{\rm t}_\odot}={(M_c\sin i)^3\over
  M^2}{\rm~M}_\odot\,, 
\ee
where $M=M_p+M_c$ is the total mass, $M_p$ is the pulsar mass, and
$M_c$ is the companion mass (all measured in M$_\odot$ units), 
and t$_\odot=G{\rm M}_\odot/c^3$ is 4.9255 $\mu$s.  The
mass function $f_p$ also equals the minimum possible mass $M_c$. 
Note that even if the difficult to measure 
inclination angle $i$ is known, both $M_p$ and $M_c$ can be inferred 
only if the mass function $f_c$ of the companion is also measurable.
This is possible in the rare case when the companion is itself a pulsar or a star with an observable
spectrum.

Binary pulsars being compact systems, several general
relativistic effects can often be observed.  These include the advance
of the periastron of the orbit 
\be
\dot\omega=3\left(\frac {2\pi}{P}\right)^{5/3}(M{\rm t}_\odot)^{2/3}
(1-e^2)^{-1}\,, 
\ee 
the combined effect of variations in the tranverse Doppler shift and
gravitational redshift (time dilation) around an elliptical orbit 
\be
\gamma=e\left(\frac{P}{2\pi}\right)^{1/3}\frac{M_c(M+M_c)}{M^{4/3}}{\rm
  t}_\odot^{2/3}\,, 
\ee 
and the orbital period decay due to the emission of gravitational radiation 
\be 
\dot P=-{192\pi\over5}\left({2\pi{\rm T}_\odot\over
P}\right)^{5/3}\left(1+{73\over24}e^2+{37\over96}e^4\right)
(1-e^2)^{-7/2}{M_pM_c\over M^{1/3}} \,.  
\ee 
In some fortunate cases, the Shapiro delay~\cite{Shapiro64}
caused by the passage of the pulsar signal
through the gravitational field of its companion can be measured. 
This general relativistic effect produces a
delay in pulse arrival time~\cite{damour86,freirewex10}
\be 
\delta_S(\phi)=2M_c{\rm
t}_\odot\ln\left[{1+e\cos\phi\over1-\sin(\omega+\phi)\sin i}\right]\,,
\ee 
where $\phi$ is the true anomaly, the angular parameter defining the
position of the pulsar in its orbit relative to the periastron.  The arrival time $\delta_S$ 
is a periodic function of $\phi$ with an amplitude
\be
\Delta_S\simeq2M_ct_\odot\left|\ln\left[\left({1+e\sin\omega\over1-e\sin\omega}\right)\left({1+\sin i\over1-\sin i}\right)\right]\right|\,.
\label{ds}
\ee 
For edge-on binaries with $\sin i\sim1$, or those which have
both large eccentricities and large magnitudes of $\sin\omega$, the amplitude $\Delta_S$ becomes very large and measurable.   
For circular orbits with $\sin~i\simeq 1$, 
Eq. ({\ref{ds}) reduces to~\cite{LP:11} 
\be
\Delta_s\simeq 4M_c t_\odot\ln \left( \frac{2}{\cos i} \right) \,,
\label{ads}
\ee
highlighting the role of the companion mass $M_c$ and the inclination angle $i$ in controlling the magnitude of $\Delta_S$. 
Table I shows a compilation of the measured Shapiro delays in which expectations from Eq.~(\ref{ads}) are compared with 
the measured values (column under Full).

%
%%%%%%%%%%%%%%%%%%%%%%%%%%%%%%%%%%%%%%%%
\begin{table}[hbt]
\begin{center}
%\begin{ruledtabular}
\begin{tabular}{|l|lll|l|ll|l|}
\hline
Pulsar  & $M_P$ & $M_c$ & $i$ & Full & Abs & RMS  & Eq. (\ref{ads}) \\ \hline
J0437-4715 & 1.76 & 0.254 & 42.42 & 4.08 & 0.25 & 0.20 & 4.1 \\
B1855+09 & 1.5 & 0.258 & 86.7 & 17.94 & 9.56 & 1.00 & 18.03 \\
J1713+0747 & 1.3 & 0.28 & 72.0 & 10.11 & 2.60 & 0.40 & 10.19 \\
J1640+2224 & Unk. & 0.15 & 84 & 8.67 & 3.94 & 1.0 & 8.71 \\
J0737-3039A & 1.3381 & 1.2489 & 88.7 & 109.64 & 68.26 & 18.00 & 110.21 \\
J1903+0327 & 1.667 & 1.029 & 77.47 & 44.56 & 14.69 & 1.00 & 44.79 \\
J1909-3744 & 1.438 & 0.2038 & 86.58 & 14.03 & 7.42 & 0.07 & 14.1 \\
J1614-2230 & 1.97 & 0.500 & 89.17 & 48.29 & 31.65 & 1.10 & 48.54 \\
J1802-2124 & 1.24 & 0.780 & 80 & 37.25 & 13.83 & 2.20 & 37.44 \\
J0348+0432 & 2.01 & 0.172 & 40.2 & 2.59 & 0.14 & 10.00 & 3.0 \\
\hline
\end{tabular}
%\end{ruledtabular}
\caption[Shapiro delays.]{Entries with numerical values are as follows.
$M_P$ {\rm and} $M_c$: Pulsar and companion masses in ${\rm M}_\odot$, 
$i$: Approximate inclination of the source in degrees, 
Full: Peak-of-cusp to bottom-of-delay Shapiro signal  amplitue in $\mu$s,  
Abs: Approximate detectable Shapiro amplitude (rest gets fits wrongly),  
RMS: Approximate rms timing residuals for the pulsar in $\mu$s, and 
Eq.~(\ref{ads}): Estimate to be compared with entries in Full. 
Table courtesy of Scott Ransom and Paul Demorest (NRAO).}
\end{center}
\end{table}
%%%%%%%%%%%%%%%%%%%%%%%%%%%%%%%%%%%%%%%%%%%%%%%%
%

Neutron star masses have also been inferred from measurements involving X-ray/Optical, double neutron star, white-dwarf-neutron star and main sequence-neutron star binaries, although not with the same accuracy characteristic of radio pulsar measurements (see \cite{LP:11,Lat12} for summaries) . 
Measured neutron star masses with 1-$\sigma$ errors can be found in the compilation of Lattimer, who maintains a contemporary table, figure and references in  http://www.stellarcollapse.org.
Recent discoveries of the $1.97\pm 0.04~{\rm M}_\odot$ pulsar in PSR J1614-2230 \cite{demorest10} and  
 $2.01\pm 0.04~{\rm M}_\odot$ pulsar in PSR J0348+0432 \cite{antoniadis13} have caused much excitement insofar as these well-measured masses severely restrict the EOS of neutron star matter.   Masses well in excess of 2 ${\rm M}_\odot$, 
 albeit with large uncertainties,  have also been reported, as e.g., $2.44^{+0.27}_{-0.27}~{\rm M}_\odot$ for 4U 1700-377, $2.39^{+0.36}_{-0.29}~{\rm M}_\odot$ for PSR B1957+20 both in X-ray binaries, and $2.74\pm 0.21~{\rm M}_\odot$ for 
 J1748-2021B in neutron star-white dwarf binaries. 

\subsection{Neutron star radii}
\label{Sec:radii}
To date, data on radii to the same level of accuracy that radio pulsar measurements on masses of neutron stars have afforded us do not exist.  Precise measurements of the mass and radius of the same neutron star would be a first and an outstanding achievement in neutron star research. Such data on several individual neutron stars would pin down the EOS of neutron star mater without recourse to models.  Recognizing the importance of radius measurements, significant efforts have been made that include observations of \\
%
%\begin{enumerate}
%\item 
1. isolated neutron stars and intermittently quiescent neutron stars that undergo accretion of matter from a companion star, and \\
%\item 
2. neutron stars that exhibit type I X-ray bursts from their surfaces.   \\
%\end{enumerate}
A brief account of the current status is provided below (see Ref. \cite{Lat12} for more details).
\vspace*{-0.25in}
\subsubsection{Isolated neutron stars}
Discovered in the all-sky search of the Rosat observatory, and thereafter investigated by the Chandra, HST and XMM observatories, there are currently 7 isolated neutron stars, referred to as the ``magnificent seven'',  from which predominantly thermal emission from the surface has been detected (see Table II).  Recent reviews, from which the data below are extracted, can be found in, e.g.,  \cite{Lat12,Kaplan08}, and references therein.
%
%%%%%%%%%%%%%%%%%%%%%%%%%%%%%%%%%%%%%%%%
\begin{table}[hbt]
\begin{center}
%\begin{ruledtabular}
\begin{tabular}{|c|c|c|c|} 
\hline
Star & $T_\infty$ & $P$ & $D$    \\ 
             & (eV) & (s) & (pc) \\
\hline
RX J0420.0-5022 & 44 & 3.45 & $\cdots$ \\
RX J0720.4-3125 & 85-95 & 8.39 & $330^{+170}_{-80}$ \\
RX J0806.4-4123 & 96 & 11.37 & $\cdots$ \\
RX J1308.8+2127 & 86 & 10.31 &$\cdots$  \\
RX J1605.3+3249 & 96 & 6.88? & $\cdots$ \\
RX J1856.5-3754 & 62 & 7.06 & $120\pm8$ \\
RX J2143.0+0654 & 102 & 9.44 & $\cdots$\\
\hline
\end{tabular}
%\end{ruledtabular}
\caption[RX J's]{Some properties of the ``magnificent seven'' isolated neutron stars. 
The temperature $T_\infty$ is inferred by spectral analysis. 
The spin period $P$ of these radio-quiet stars is inferred from X-ray pulsations. Only in one case is  the distance $D$ to the star well known.}
\end{center}
\end{table}
%%%%%%%%%%%%%%%%%%%%%%%%%%%%%%%%%%%%%%%%%%%%%%%%
%

The observed flux (in all cases in X-rays, and when the star is nearby enough in optical as well) is generally fit using 
\be
F = 4\pi \sigma~ T_\infty^4 \left(\frac {R_\infty}{D} \right)^2  \qquad {\rm and} \qquad 
R_\infty = R~\left( 1 - \frac {2GM}{c^2R} \right)^{-1/2} \,,
\ee
where $\sigma$ is Boltzmann's constant, $T_\infty = T \left[ 1-2GM/(c^2R)\right]^{1/2}$ is an effective temperature that fits the data well, and, the so-called ``radiation radius'' $R_\infty$ is related to the mass $M$ and insitu radius $R$ of the star as indicated above .  
The subscript $\infty$ in the above relations refers to an observer situated at a far distance from the source. The distance to the star, $D$, is generally beset with large uncertainties unless determined through parallax and proper motion measurements (as in the case of RX J1856.5-3754 \cite{Walter96,Walter02}).  Using $T_\infty$, $R_\infty$ and the surface redshift parameter $z =  \left[ 1-2GM/(c^2R)\right]^{-1/2} - 1$ as parameters in the spectral analysis, 
the radius and mass  of the star can be determined through
\be
R = R_\infty ~(1+z)^{-1} \qquad {\rm and} \qquad 
\frac {M}{\rm {M}_\odot} = \frac {c^2R}{2G{\rm M}_\odot} ~ \left[1-(1+z)^{-2}\right] \,. 
\ee
Real life, however, intervenes to destroy the simplicity of the above procedure.  A neutron star is not a perfect blackbody (as is implicit in the above expressions). The star' s unknown atmospheric composition, strength and structure of the magnetic field, interstellar hydrogen absorption, etc., all of which shape the observed spectra, must be accounted for. A case in point is inferences drawn from the best studied case of the nearest known isolated neutron star RX J1856-3754.  Depending upon the atmospheric model used, the inferred masses vary significantly, although the radii are similar (see Table III).   Non-magnetic heavy element atmospheres \cite{Pons02,Walter04} predict spectral features that are not observed. Following indications of a surface magnetic field of $B_S \sim 5\times 10^{12}$ G, magnetized and condensed surfaces have been investigated \cite{Burwitz03,Ho07}, but require trace elements of H with a  finely tuned mass (origin unknown) to adequately fit the data. Despite much promise, reliable extractions of $M$ and $R$ from isolated neutron stars awaits further developments in the treatment of atmospheres, and additional data. The magnificence of the seven is yet to be realized!

%
%%%%%%%%%%%%%%%%%%%%%%%%%%%%%%%%%%%%%%%%
\begin{table}[hbt]
\begin{center}
%\begin{ruledtabular}
\begin{tabular}{|c|c|c|c|c|c| }
\hline
 $R_\infty$ (km) & $z$ & $R$ (km) & $M~({\rm M}_\odot) $ & Atmospheric model & Ref.   \\ 
\hline
 $16.1\pm1.8$ & $0.37\pm0.03$ & $11.7\pm1.3$ & $1.86\pm0.23$ & Non-magnetic heavy elements & \cite{Pons02} \\  
 $\simeq 15.8$ & $\simeq 0.3$ & $\simeq 12.2$ & $\simeq 1.68$ & Non-magnetic heavy elements & \cite{Walter04} \\ 
 $>13$ & $\cdots$  & $\cdots$ &$\cdots$ &Condensed magnetized surface & \cite{Burwitz03} \\
 $14.6\pm1$ & $\simeq0.22$ & $11.9\pm0.8$ & $1.33\pm0.09$ & Condensed magnetized surface; trace H & \cite{Ho07} \\
\hline
\end{tabular}
%\end{ruledtabular}
\caption[RX J1856-3754]{Inferred mass and radius of the isolated neutron star RX J1956-3754 from different models of atmospheres using data from Rosat, HST, Chandra and XMM observatories.}   
\end{center}
\end{table}
%%%%%%%%%%%%%%%%%%%%%%%%%%%%%%%%%%%%%%%%%%%%%%%%
%
\vspace*{-0.25in}
\subsubsection{Quiescent neutron stars}
Between episodes of intermittent accretion from a companion star, many neutron stars are known to go through long periods of quiescence. Accretion of matter induces compression of matter in the crust of a neutron star triggering pycno-nuclear reactions that release energy \cite{Haensel90} which heats the crust. During the quiescent periods, the heated crust cools and radiates detectable X-rays \cite{Brown98}.  
Due to the lack of evidence for significant magnetic fields, such as pulsations or cyclotron frequencies, 
the observed spectra are generally fitted with non-magnetic H atmospheres that are well understood.  Models to
infer the apparent angular emitting area and the surface gravity \cite{Heinke06,Webb07,Guillot11} have resulted in 
probability distributions of $M$ and $R$,  four of which in globular clusters M13, X7, $\omega$~Cen and U24 are shown in Figure 10 of Ref. \cite{Lat12}, courtesy of A. W Steiner.  The results are such that wide ranges in $M$ and $R$ values are permitted for each of the four cases considered.  
 
\vspace*{-0.25in}
\subsubsection{Type I X-ray bursts undergoing photospheric radius expansion}
Subsequent to accretion, the envelope of a neutron star can become thermally unstable to He or H ignition which leads to a thermonuclear explosion observed as an x-ray burst with a rapid rise time ($\sim$ 1 s) followed by a cooling stage lasting to $\sim$ 10-100 s \cite{Stroh04}.  For sufficiently luminous bursts, the surface layers of the neutron star and the photosphere are driven outward to larger radii by the radiation pressure.  The flux at the photosphere can approach or even exceed the Eddington value for which the radiation pressure balances gravity. The bursters EXO 1745-248, 4U 1608-522, 4U 1820-30 and KS 1731 have been modeled in Refs. \cite{Ozel12,Steiner10,Steiner13,Suleimanov11} to infer masses and radii of neutron stars.  The key physical parameters of these models are the opacity of the lifted material, the effective blackbody temperature when the lifted material falls down to the 
surface after expansion (touchdown), the color correction factor that accounts for effects of the atmosphere in distorting the inferred temperature, possible models of atmospheres, and whether or not the photosphere radius is equal to or larger than the radius of the neutron star.  The inferred values of radii have ranged from 8-10 km \cite{Ozel12}, 11-13 km \cite{Steiner10,Steiner13} and in excess of 14 km \cite{Suleimanov11}. The situation is far from settled, although firm beliefs are held by each group of analyzers.

\section{NEUTRON STAR STRUCTURE AND THE EQUATION OF STATE}
\label{Sec:Structure}
In old and cold neutron stars, matter is in weak-interaction equilibrium and charge neutral. 
It suffices to choose two independent chemical potentials $\mu_n$ and $\mu_e$ to characterize the prevalent conditions (neutrinos with their long mean free paths leave the star; when trapped, their chemical potentials have to be counted). For example, $\mu_n -\mu_p= \mu_e = \mu_\mu$ (energy conservation) and $n_p= n_{e} + n_{\mu}$ (charge neurtrality) in nucleons only matter, where the subscripts $n,~p~,e~{\rm and}~\mu$ denote neutrons, protons, electrons, and muons, respectively. When other baryons besides nucleons, or mesons or quarks, are present similar relations are straightforwardly deduced (see, e.g., \cite{Prak96}). With the composition of matter thus determined,  the relation between pressure $p$ and energy density $\epsilon$, or the equation of state (EOS), can be determined using models of strong interacting matter (not certain yet) and leptons (noninteracting contributions suffice as those from interactions are negligibly small). 

In hydrostatic equilibrium, the structure of a spherically symmetric neutron star is determined by the Tolman-Oppenheimer-Volkov (TOV) equations \cite{TOV1,TOV2}
\be
\frac {dp(r)}{dr} = - \frac {G}{c^2} \frac {[p(r)+\epsilon(r)]~[m(r)+4\pi r^3 p(r)/c^2]} {[(r-2Gm(r)/c^2]} \quad {\rm and} \quad 
\frac {dm(r)}{dr}  = 4\pi r^2~\frac {\epsilon(r)}{c^2}\,, 
\label{TOV}
\ee
where $m(r)$ is the enclosed mass at radius $r$.  The gravitational and baryon masses are given by
\be
M_Gc^2 = \int_0^R dr 4\pi r^2 \epsilon(r) \quad {\rm and} \quad M_bc^2 = m_b~\int_0^R dr ~4\pi r^2 n(r) 
\left[ 1 - \frac{2Gm(r)}{c^2r} \right]^{-1/2} \,,  
\ee
where $m_b$ is the baryonic mass and $n(r)$ is the baryon number density.   
With the EOS $p=p(\epsilon)$ as input (chiefly from strong interaction theory), the structure of the star is determined by specifying a central pressure $p_c=p(\epsilon_c)$ at $r=0$ and integrating the above coupled differential equations out to the star surface at $r=R$ where $p(r=R)=0$. The binding energy of the star is then B.E. = $(M_b-M_G)c^2$.  
The results allow us to map out predictions for $M$ vs. $R$, $M$ vs $n_c$, B.E. vs $M_G$, etc. (see, e.g., \cite{Prak96}).

\subsection{What can be said on the basis of masses alone?}
The implications of masses  in excess of  2~${\rm M}_\odot$ are illustrated in Fig. \ref{ultimate} where  the relation between the maximum mass and the central energy density  $\epsilon_c$ (and 
baryon number density $n_c$) resulting from various proposed EOS's are shown (cf. Refs. \cite{LP:05,LP:11} for details).  
 
 %
%%%%%%%%%%%%%%%%%%%%%%%%%%%%%%%%%%%%%%%%
%
\begin{figure}[thb]
\vskip -1cm
\begin{center}
{\includegraphics[width=300pt,angle=90]{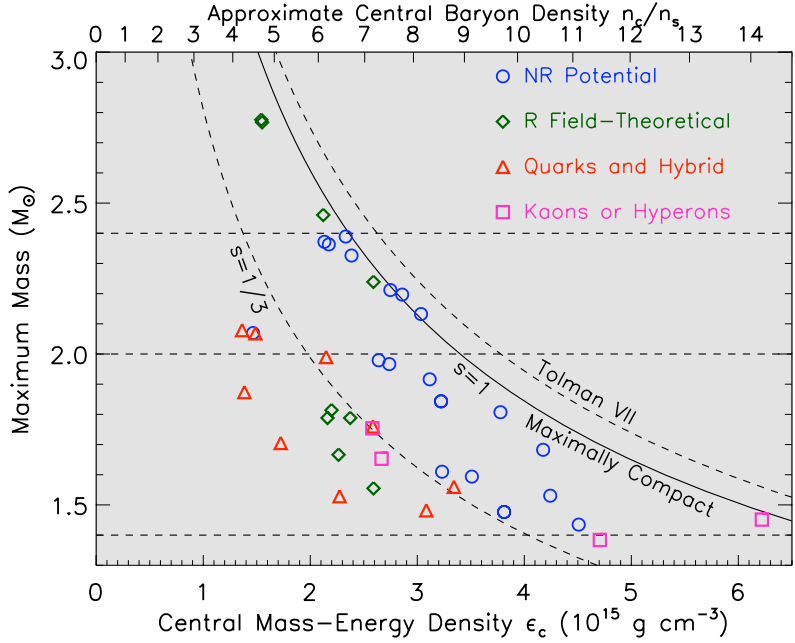}}
\end{center}
\vskip -1.5cm
\caption{(Color online) Maximum mass versus central mass-energy density (bottom $x$-axis) and central baryon density (top $x$-axis) 
for the maximally compact EOS in Eq. (\ref{compeos}). 
The curve labelled $s=1/3$ corresponds to  $p=(\epsilon-\epsilon_0)/3$ characteristic of commonly used quark matter EOSs.  Results of Tolman VII solution \cite{Tolman39}  with  
$\epsilon=\epsilon_c(1-(r/R)^2)$ and for various model calculations of neutron star matter - see inset for legends - are as shown. Figure adapted from Ref. \cite{LP:05}.}
\label{ultimate}
\end{figure}
%
%%%%%%%%%%%%%%%%%%%%%%%%%%%%%%%%%%%%%%%%
%
%\cite{Tolman39} 
%Ref. \cite{LP05}.}

\noindent Fig. \ref{ultimate} also shows results from useful schematic EOS's that provide bounds. 
 The most compact and massive configurations are obtained when the low-density EOS is ``soft'' and the high-density EOS is ``stiff'' \cite{Haensel89,Koranda97}. 
 Using limiting forms in both cases, the maximally compact EOS is therefore given by the pressure ($p$) vs. energy density $(\epsilon)$ relation
 \be
 p = 0 \quad {\rm for} \quad \epsilon < \epsilon_0\,; \quad p = \epsilon - \epsilon_0 \quad {\rm for} \quad \epsilon > \epsilon_0\,.
 \label{compeos}
 \ee
Above, the stiff EOS is at the causal limit as $dp/d\epsilon=(c_s/c)^2=1$, where $c_s$ is the adiabatic speed of sound. This EOS has a single parameter $\epsilon_0$ and the structure (TOV) equations scale with it according to \cite{Witten84}
 \be
 \epsilon \propto \epsilon_0\,, \qquad p \propto \epsilon_0\,, \qquad m \propto \epsilon_0^{-1/2}, \qquad {\rm and} \qquad 
 r \propto \epsilon_0^{-1/2} \,,
 \ee
 where $m$ is the star's enclosed mass and $r$ its radius. Employing these scaling relations,  
 the compactness ratio $(GM_{max}/R_{max}c^2)$ is smallest when \cite{Koranda97,LP:11}
 \be
 M_{max} = 4.09~ (\epsilon_s/\epsilon_0)^{1/2} ~{\rm M}_\odot\,, \quad 
 R_{max} = 17.07~ (\epsilon_s/\epsilon_0)^{1/2}~{\rm km}\,,  \quad {\rm and} \quad
 BE_{max} = 0.34~M_{max}c^2  
 \label{maxMR}
 \ee  
 where $\epsilon_s \simeq 150~{\rm MeV~fm}^{-3}$ is the energy density at the nuclear saturation density of $n_0=0.16~{\rm fm}^{-3}$. If the EOS is deemed known up to $\epsilon_0\sim 2\epsilon_s$, the maximally compact EOS yields $M_{max} \sim 3~{\rm M}_\odot$.     The upper limits on the corresponding thermodynamic variables are \cite{Koranda97,LP:11}:
 \be
 \epsilon_{max} = 3.034~\epsilon_0\,, \quad p_{max} = 2.034~\epsilon_0\,, \quad \mu_{max} = 2.251~\mu_0\,, 
 \quad {\rm and} \quad n_{max} = 2.251~(\epsilon_0/\mu_0)\,,
\label{maxthermo}
\ee
 where $\mu_0\simeq 930$ MeV is the mass-energy of iron nuclei per baryon in a star with a normal crust. Combining Eqs. 
 (\ref{maxMR}) and (\ref{maxthermo}), we arrive at the result \cite{LP:11}
 \be
 \epsilon_{max} \leq 50.8~ \epsilon_s ~({\rm M}_\odot/M_{max})^2\,, 
 \label{Ultimate}
 \ee
 a relation that enables us to appreciate the impact of the maximum mass of a neutron star on the ultimate energy density of cold observable matter.  If the largest measured mass represents the true neutron star maximum mass, it sets upper limits on the central energy density, pressure, baryon number density and chemical potential. In the case of the 1.97 M$_\odot$, these limits turn out to be
 \be
 \epsilon_{max} < 1.97~{\rm GeV~fm}^{-3}, \quad p_{max} < 1.32~{\rm GeV~fm^{-3}}, \quad 
 n_{max} < 1.56~{\rm fm^{-3}}, \quad \mu_{max}  < 2.1~{\rm GeV}\,.  
 \ee
Substantial reductions in the energy density and baryon number density occur if a well-measured mass exceeds  2.0 M$_\odot$, as illustrated by the case of a 2.4 M$_\odot$ in Fig. \ref{ultimate}. 

\subsection{Self-bound quark stars}
An analysis for the general EOS $p=s(\epsilon-\epsilon_0)$ can be found in Ref. \cite{LP:05} for various values of $s$. The case $s=1/3$ and $\epsilon_0=4B$ corresponds to the MIT bag model quark matter EOS with $B$ being the bag constant.  
In this case, maximally compact configurations are characterized by
 \ba
 M_{max} &=& 2.48~ (\epsilon_s/\epsilon_0)^{1/2} ~{\rm M}_\odot \quad 
 R_{max} = 13.56~ (\epsilon_s/\epsilon_0)^{1/2}~{\rm km}\,,  \quad {\rm and} \quad 
 BE_{max} = 0.21~M_{max}c^2
 \nonumber \\
 \label{maxMRQM}
\epsilon_{max} &\simeq& 30\left(\frac{{\rm M_\odot}}{M_{max}}\right)^2~ \epsilon_s\, , 
\quad p_{max}  \simeq 7.9\left(\frac{{\rm M_\odot}}{M_{max}}\right)^2~ \epsilon_s\, , 
 \quad n_{B,max}   \simeq 27\left(\frac{{\rm M_\odot}}{M_{max}}\right)^2~ n_s\, , 
  \quad {\rm and} \nonumber \\
 \quad \mu_{B,max} &\simeq& 1.46~{\rm GeV}\,,  
\label{maxthermoQM}
 \ea  
where a value of $\mu_0=930$ MeV was used as self-bound quark stars are expected to have a very thin crust (that does not affect $M$ and $R$ significantly) of normal matter. The $M_{max}$ versus $\epsilon_c$ curve for $s=1/3$ shown in Fig. \ref{ultimate} lies a factor of $\sim 0.6$ below the $s=1$ curve. Effects of adding QCD corrections, finite strange quark mass and CFL gaps makes the EOS more attractive and less compact \cite{LP:11}. Noteworthy is the relatively low value of the baryon chemical potential (1.46 GeV), which calls for non-perturbative treatments of quark matter.

\subsection{Hybrid stars containing quark mater}
\vskip -0.25cm

%%%%%%%%%%%%%%%%%%%%%%%%%%%%%%%%%%%%%%%%
%
\begin{figure}[thb]
\vskip -0.25cm
%\begin{center}
{\includegraphics[width=220pt,angle=0]{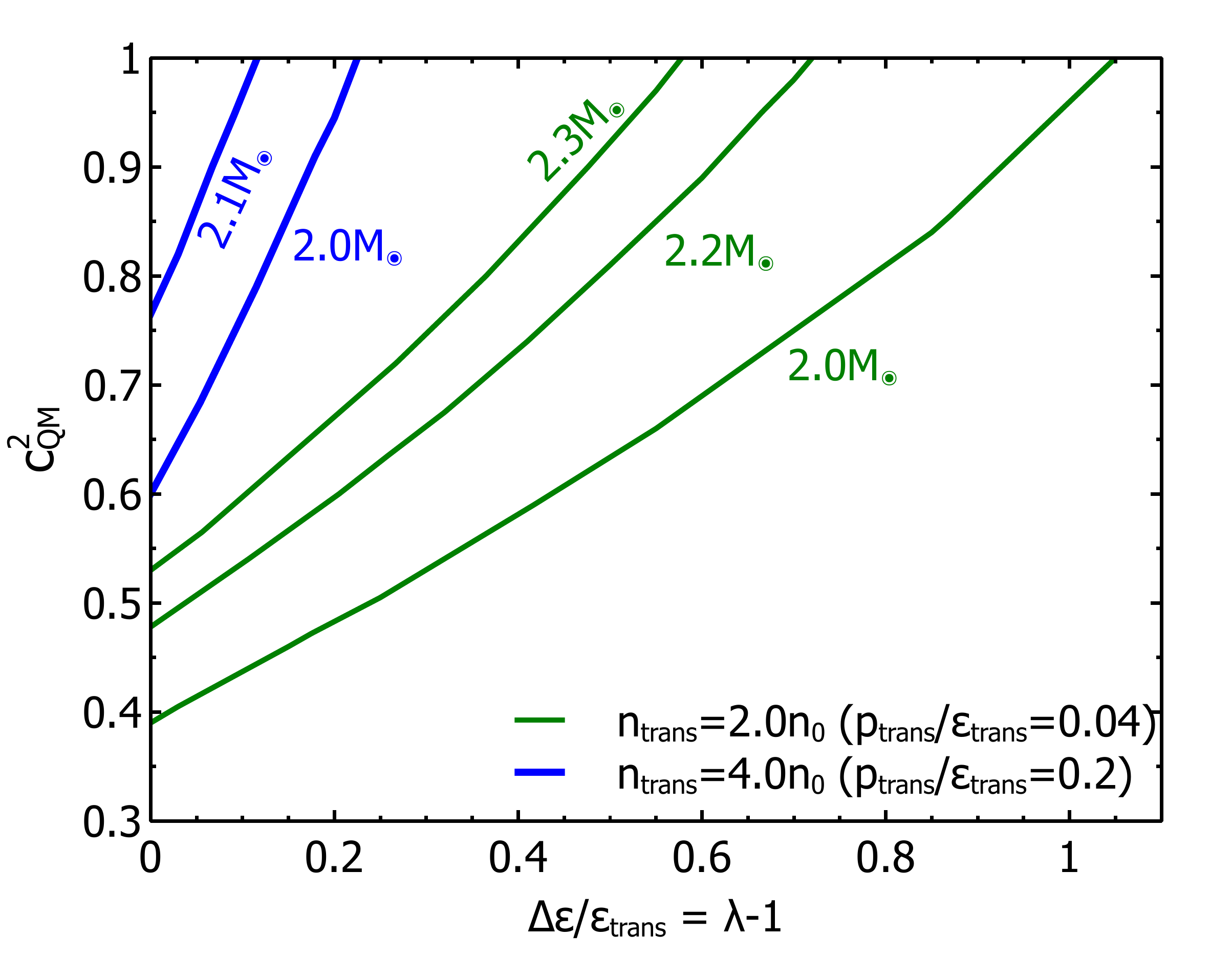}}
{\includegraphics[width=220pt,angle=0]{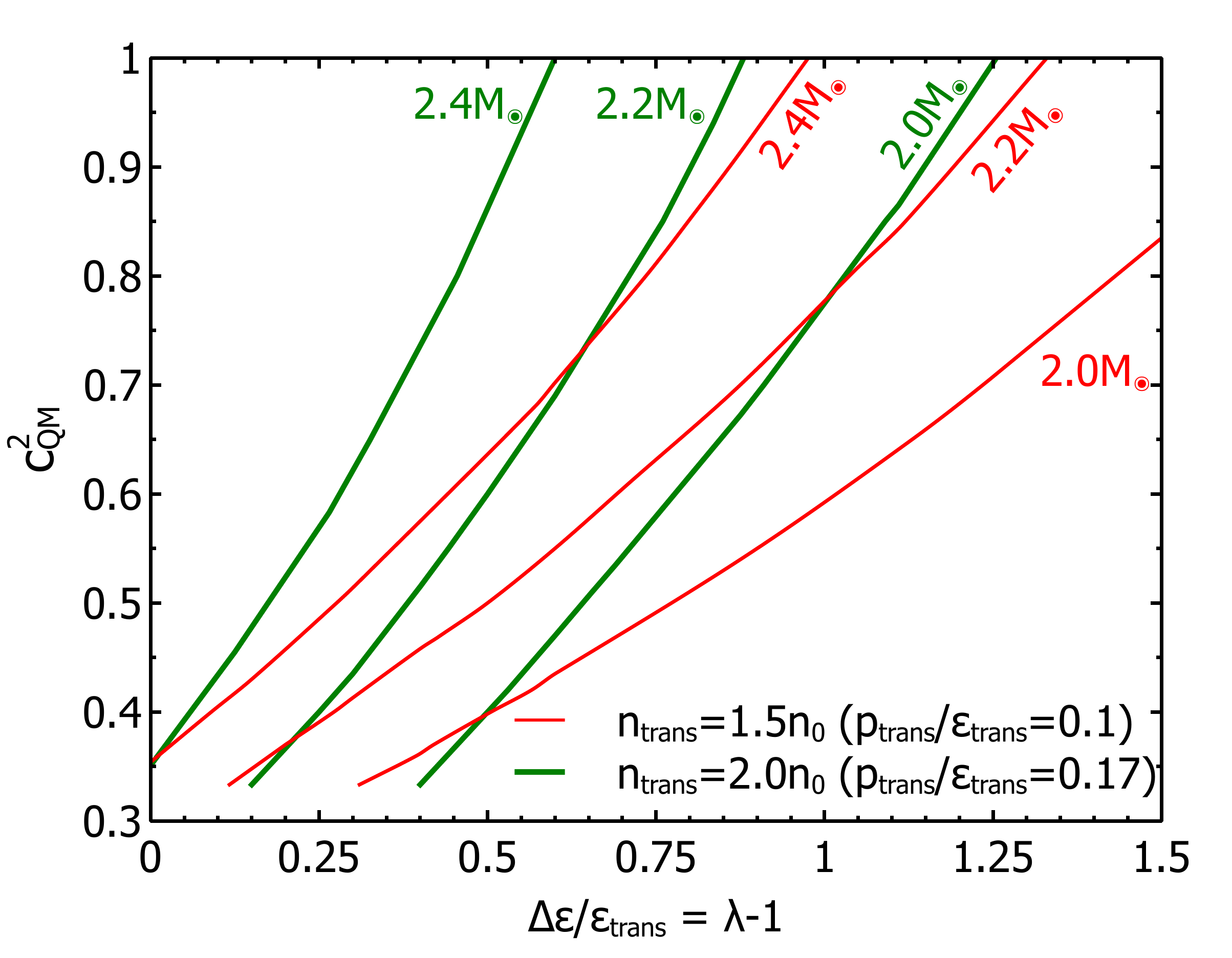}}
%\end{center}
%\vskip -2.0cm
\caption{(Color online) Mass of the heaviest hybrid star as a function of quark matter EOS parameters $p_{trans}/\epsilon_{trans}$, $c_{QM}^2$, and $\Delta\epsilon/\epsilon_{trans}$ 
for HLPS (left panel) and NL3 (right panel) nuclear matter. The thin (red), medium (green) and thick (blue) lines are for nuclear to quark transition at 
$n_{trans}=1.5n_0,~2n_0$ and $4n_0$, respectively. Figure adapted from Ref. \cite{Alford13}.}
\label{maxmasses}
\end{figure}
%
%%%%%%%%%%%%%%%%%%%%%%%%%%%%%%%%%%%%%%%%
Recently, Ref. \cite{Alford13} examined hybrid stars assuming a single first-order phase transition between nuclear and quark matter, with a sharp interface between the quark matter core and nuclear matter mantle.  To establish generic conditions for stable hybrid stars, the EOS of dense matter was taken to be
\be
\epsilon(p) = \left\{\!
\begin{array}{ll}
\epsilon_{\rm NM}(p) & \quad p<p_{trans}\\
\epsilon_{\rm NM}(p_{trans})+\Delta\epsilon+c_{\rm QM}^{-2} (p-p_{trans}) & \quad p>p_{trans}
\end{array}
\right.
\label{eqn:EoSqm1}
\ee
where $\epsilon_{\rm NM}(p)$ is the nuclear matter equation of state, $\Delta\epsilon$ is the discontinuity in energy density $\epsilon$ at the transition pressure $p_{trans}$, and $c_{QM}^2$ is the squared speed of sound of quark matter taken to be constant with density (as in a classical ideal gas) but varied in the range 1/3 (roughly characteristic of perturbative quark matter) to 1 (causal limit).   Two illustrative examples for $\epsilon_{{\rm NM}}(p)$: a 
relativistic mean field model labelled NL3 \cite{Shen:2011kr} 
and a non-relativistic potential model labelled HLPS,
corresponding to ``EoS1'' in Ref.~\cite{Hebeler:2010jx} are shown in Fig. \ref{maxmasses}.  Insofar as HLPS is softer than NL3, these EOS's provide a contrast  at low density. 
The principal finding is  that   it is possible to get  hybrid stars in excess of 2 M$_\odot$ for reasonable parameters of the quark matter EOS. 
The requirements are not-too-high transition density ($n\sim 2n_0$), low enough energy density discontinuity $\Delta\epsilon < 0.5~\epsilon_{trans}$, and high enough speed of sound $c_{QM}^2 \geq 0.4$.
 It is worthwhile to note that perturbative treatments are characterized by $c_{QM}^2\simeq1/3$, and a value of $c_{QM}^2$ well above 1/3 is an indication that quark matter is strongly coupled. Clearly, non-perturbative treatments of quark matter are indicated.

In summary, larger the observed neutron star mass,  larger is the challenge for theory to come up with an EOS that can support it. The lower the mass, larger is the challenge to devise  a stellar evolutionary scenario to form such a low mass  given the current paradigm of core collapse supernovae \cite{Lat12}.  Clearly, the maximum and minimum masses  of neutron stars are of paramount importance to nuclear/particle theory, astrophysics, and cosmology.      

\section{Toward a model-independent EOS of neutron star matter}
\vskip-0.25cm
%%%%%%%%%%%%%%%%%%%%%%%%%%%%%%%%%%%%%%%%
%
\begin{figure}[thb]
\vskip -0.25cm
{\includegraphics[width=220pt,angle=0]{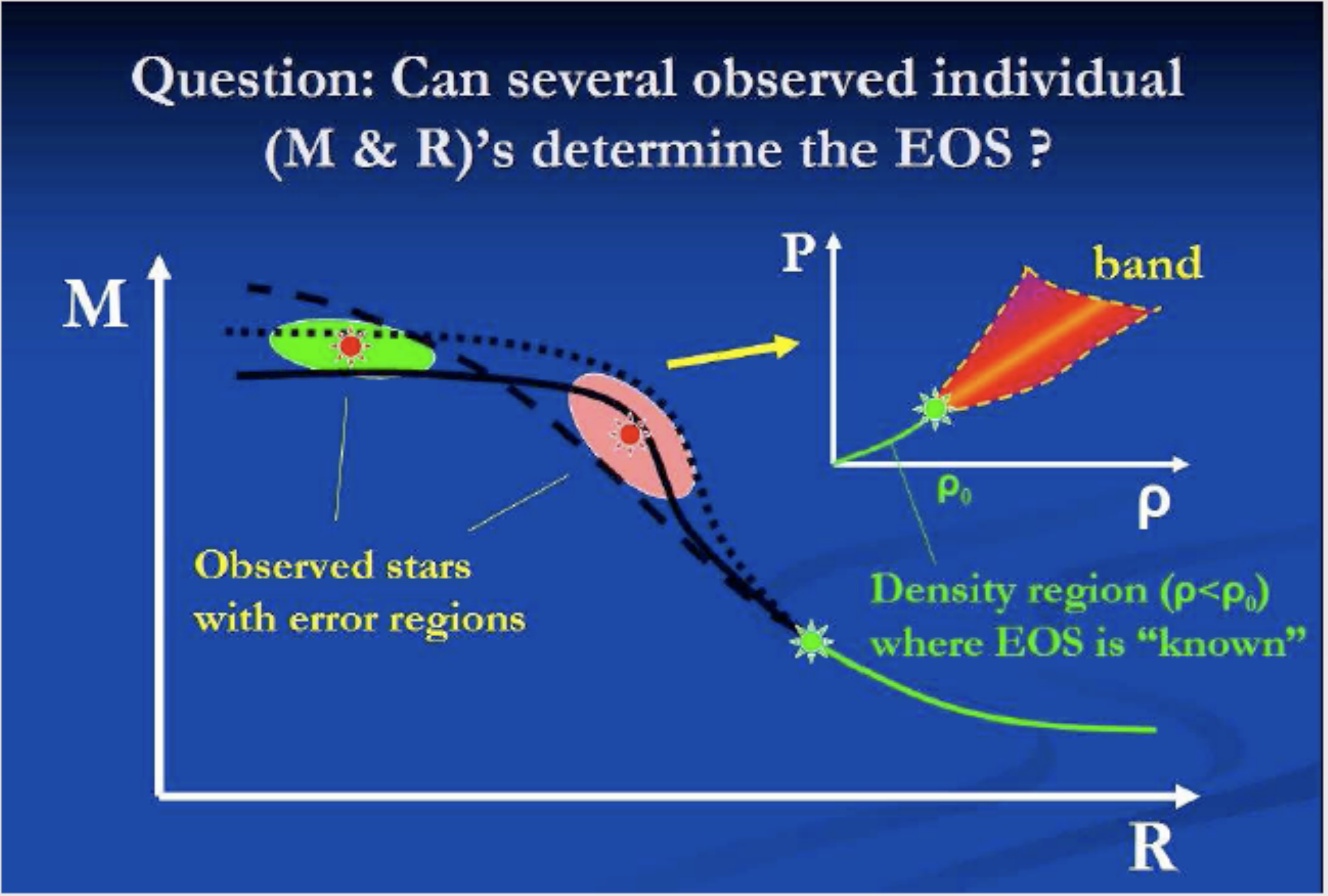}}
{\includegraphics[width=220pt,height=150pt,angle=0]{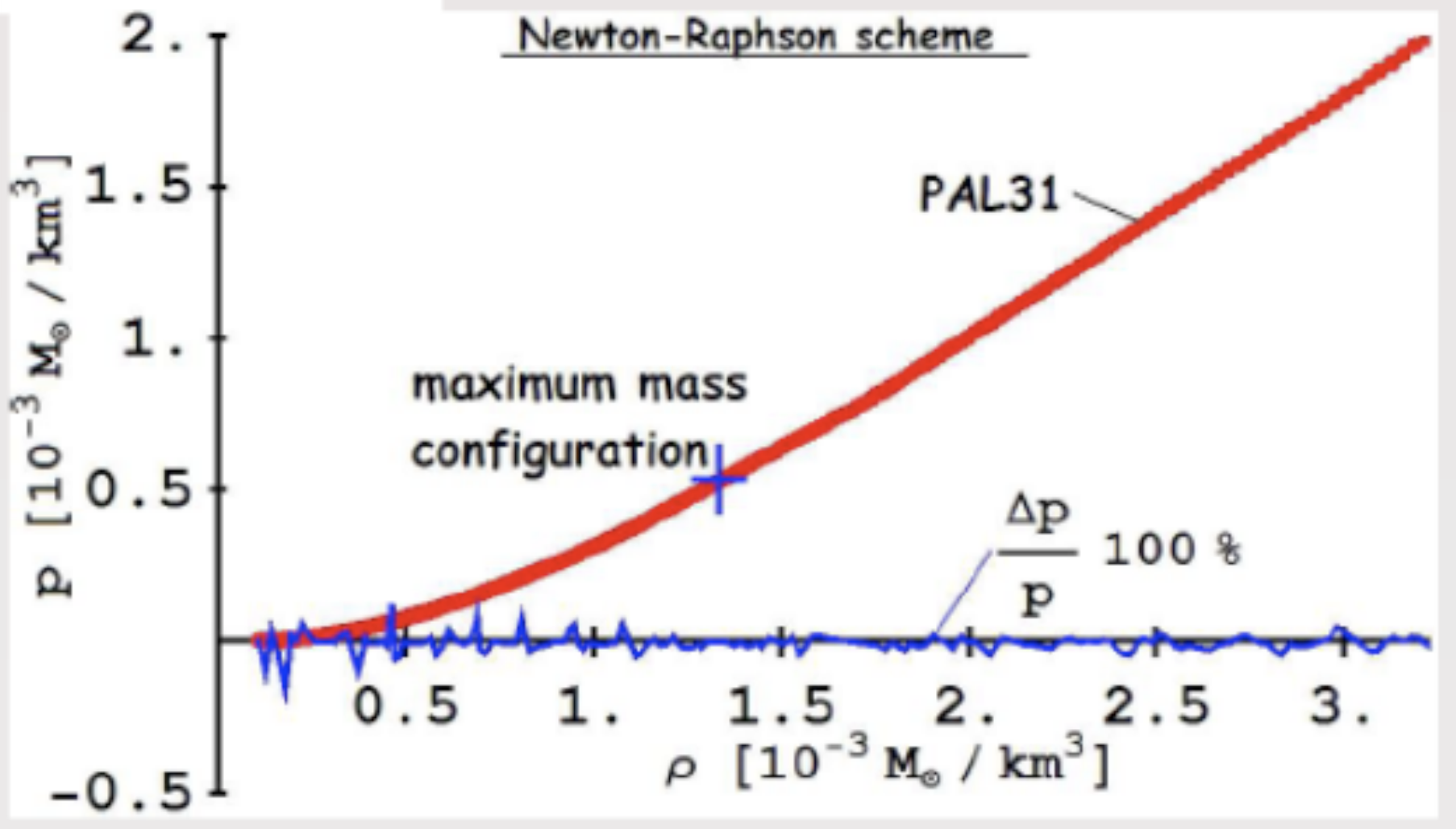}}
\caption{(Color online) Deconstructing a neutron star with a physically motivated nucleonic EOS. Left panel figure courtesy Postnikov. Right panel figure adapted from Ref. \cite{Prakash09b}.} 
%\cite{Prakash09}.}
\label{deconstruct}
\end{figure}
%
%%%%%%%%%%%%%%%%%%%%%%%%%%%%%%%%%%%%%%%%
Accurately measured masses and radii of several individual neutron stars can uniquely determine the dense matter EOS in a model-independent manner. The method, developed by Lindblom \cite{Lind:92}, exploits the one-to-one correspondence between an EOS and the $M-R$ curve generated using the TOV equations, Eq. (\ref{TOV}), rewritten as \cite{Prakash09b,Postnikov10}
\be
\frac {dr^2}{dh} = -2r^2 \frac {r-2m}{m+4\pi r^3P} \qquad {\rm and} \qquad \frac {dm}{dh} = -4\pi r^3\rho \frac {r-2m}{m+4\pi r^3P}\,,
\label{rtov}
\ee
where the pressure $p(h)$ - mas-energy density $\rho(h)$ relation constitutes the EOS. Above, the variable $h$ is defined through $dh=dp/(p+\rho(p))$. The advantages of this reformulation are that the enclosed mass $m$ and radius $r$ are now dependent variables, and $h$ is finite both at the center and surface of the star. The deconstruction procedure begins with a known EOS up to a certain density, taking small increments in mass and radius, and adopting an iterative scheme to reach the new known mass and radius. Alternatively, one can solve Eqs. (\ref{rtov}) from the center to the surface with an assumed form of the EOS using a Newton-Raphson scheme to obtain the known mass and radius. 
The right panel of Fig. \ref{deconstruct} shows results of deconstruction (from the latter scheme)
when proxy masses and radii are used from the EOS of PAL31 \cite{PAL}. Both schemes yield results to hundredths of percent accuracy. 
The number of simultaneous mass and radius measurements along with  their inherent errors will determine the accuracy with which the EOS can be determined. 
Using the currently available estimates, Steiner et al. \cite{Steiner10} have arrived at  probability distributions for pressure vs.  energy density using the $M-R$ probability distributions through a Bayesian analysis assuming a parametrized EOS. 
Theory being in place, several acurate measurements offer the promise to pin down the 
EOS of neutron star matter model-independently.

\section{Many Facets of Neutron Stars}
\label{facets}
Pulsar glitches (discontinuous decreases in rotational periods), intermittent X-ray bursts, flares in magnetars with magnetic fields  as large as $10^{15}$ G, quasi-periodic oscillations, etc., make neutron stars fascinating objects to study. Multi-wavelength photon observations have shed light on the long-term thermal evolution of neutron stars shedding light on neutrino emitting processes from their  constituents. For example, the observed surface temperature ($\sim 2\times 10^6$ K) of the 330 year old neutron star in Cassiopeia A  has confirmed the occurrence of neutron superfluidity in the dense interiors of neutron stars \cite{Page14}.  
The post-accretion thermal radiation (in X-rays) from several neutron stars has not only confirmed the theoretical prediction that neutron stars have crusts, but are also beginning to reveal the elastic and transport properties of crystalline structures in neutron star surfaces. Although much has been learned, several questions remain some of which are mentioned below.

\section{Unresolved Issues}
\label{questions}

Many longstanding questions and new ones raised by recent discoveries require answers.
\begin{enumerate}
\item 
What are the maximum and minimum masses of neutron stars?  The former  has implications for the minimum mass of a black hole (and the total number of stellar-mass black holes in our Universe), the progenitor mass,  and the EOS of dense matter. The minimum  mass raises questions about its formation through stellar evolution. 
\item 
What is the radius of a neutron star whose mass is accurately measured? 
{\it Precise measurements of masses and radii for several  individual stars would pin down the EOS without recourse to models}. 
\item
What phases are there in the phase diagram of dense matter at low temperatures? How do we use neutron star observations to learn about those phases, particularly those containing quark matter? 
\item
What limits the spin frequencies of milli-second pulsars and why? Can r-modes coupled with the presence of quark matter and its bulk viscosity be the clue solve this mystery?   
\item 
What are neutron star cooling curves telling us? Superfluidity attenuates cooling under most conditions, while exotica  (e.g., hyperons, quarks) hasten it. 
\item
Flares associated with magnetars continue to baffle us.   What is the microscopic origin of such strong surface magnetic fields and what are their magnitudes in the interiors? 
\item
What is the nature of absorption features detected from isolated neutron stars? 
\item 
What precisely controls the  durations of X-ray bursts and of inter-bursts? 
\item 
Is unstable burning of Carbon (C) the real cause of super bursts? Can the condition for igniting such burning  be met with our understanding of the C-C fusion? 
\item
Is there real evidence for enhanced neutrino cooling in high mass neutron stars? 
\item
Why do glitches occur? What triggers the coupling of the superfluid to the crust for less than a minute? What are the relevant dissipative processes? 
\item
How does one link the microphysics of transport, heat flow, superfluidity, viscosity, vortices/flux tubes to average macro-modes in neutron star phenomenology? 
\end{enumerate}
Addressing these questions requires  concerted efforts from astrophysical observations, laboratory experiments and  associated theory. Efforts in these directions include proposals for observatories such as ``The Large Observatory for X-ray Timing or LOFT'' (see 
http://sci.esa.int/loft/53447-loft-yellow-book/ [sci.esa.int] for extensive references), and, experiments with extremely neutron-rich beams at rare isotope facilities around the world.

\section*{ACKNOWLEDGEMENTS}
In addition to a large number of pals and helpful researchers who have  tutored me, 
I thank Mark Alford, Sophia Han, James Lattimer and Sergey Postnikov with great pleasure.
This research was supported from the US DOE under Grant  No. DE-FG02-93ER-40756.
 
\newpage

\bibliographystyle{h-physrev3}
\bibliography{mp_references}

\end{document}